\documentclass{desyproc}

\begin{document}
\title{Problems of determination of $\sigma_{tot}$ at the LHC}

\author{{\slshape Oleg Selyugin$^1$}\\[1ex]
$^1$BLTPh, JINR,  Dubna, Russia}

\contribID{smith\_Selyugin}


\acronym{EDS'09} 

\maketitle

\begin{abstract}
The analysis of the procedure of the determining
      the parameters of the hadron elastic scattering amplitude at
     high energy is presented. The exponential and non-exponential
     form of the scattering amplitude are taken into account.
     Especially,   the
     impact of the real parts of the hadron scattering
     amplitude and electromagnetic hadron interaction
     on the determination of the total cross section is examined.
\end{abstract}

     The experimental data obtained by the TOTEM collaboration at 7 TeV
     \cite{TOTEM-11a}  do not coincide with  the predictions of all theoretical models
     \cite{Rev-LHC}.
  The procedure  of the extracting the parameters of the scattering amplitude
  from the  data of the differential cross sections assumes  some theoretical assumption and approximations \cite{rho-pap}.
       In \cite{CS-PRL09}, as an example, it was shown that the saturation regime, which can  occur at the LHC energies,
       changes the behavior of the slope of the differential cross sections at small momentum transfer.
       As a result, the differential cross section cannot be described by  a simple exponential form with
       the constant slope.
       Another example,
       in the $p\bar{p}$ scattering at $SP\bar{P}S$ 
       there are two different measures of the
       size of $\rho$:  
       $\rho=0.24$  and 
       $\rho=0.139$. 
        More careful analysis gave $\rho=0.19$ for the data of the UA4 Collaboration \cite{Sel-YF92} and  $\rho=0.16$  for the data      of the  UA4/2 Collaboration \cite{Sel-UA42}.

         In our talk, we present  some analysis of the new experimental data obtained by the
                TOTEM Collaboration at $\sqrt{s} = 7 \ $TeV at small momentum transfer.
        In all cases, we take  into account all 5 spiral electromagnetic amplitudes
        and take into account the Coulomb-hadron interference phase \cite{selmp1}.
The hadron spin non-flip amplitude was chosen in the form
with the possible non-exponential form
    \begin{eqnarray}
 F(s,t)= (i+\rho)\frac{\sigma_{tot}}{4 k \pi } e^{[B/2 \ t \ +C /2 \ (\sqrt{4\mu^2-t}-2\mu)] ) }
 \label{fh}
  \end{eqnarray}
 where $k=0.38938 \ $ mb/GeV$^{2}$,  $t=-q^2$, and $C$ GeV$^{-1}$ is some coefficient which
 reflects  some additional part of the slope.
 We will examine the set of the TOTEM data at small $t$ with   $N=47$  points and $-t_{max} \ = \ 0.112$ GeV$^2$.
     The whole set ($  N=86 $  points and $-t_{max} \ = \ 0.3 \ $GeV$^2$)  will be examined only as an example.
     This interval of momentum transfer is large and the imaginary part of the scattering amplitude  has some
     complicated form. In all our calculations we used only statistical errors.

 Let us make the fit of the differential cross sections with the hadronic amplitude in form (\ref{fh})
    and  not take into account the electromagnetic interactions.
     We obtain the large  $\sum_{i=1}^{N} \chi^{2}_{i}$ and the size of $\sigma_{tot}= 98.7$ mb. That is practically the same, as was obtained by the TOTEM Collaboration. The result does not feel  the size of $\rho$.
 Now let us make the same fit but include the electromagnetic part of the elastic scattering amplitude.
 The first 4 rows of Table 1  present the fit without the additional part of the
         slope.  In this case, the influence of the size of $\rho$ is visible and we can make the fit taking
         $\rho$ as free parameters (the row 4). We obtained minimum $\chi^2$  with the negative size of $\rho$.
         It is essentially far away from the TOTEM analysis and the predictions of the COMPETE Collaboration.

\begin{wraptable}{r}{0.55\textwidth}
\centerline{\begin{tabular}{|c|c|c|c|}
\hline
 $\sum \chi^{2}_{i}$ & $\rho$  & $C$ & $\sigma_{tot}$, mb  \\\hline
  & & & \\
   87.1 & $0.2 $fixed      &  $0.fix $  & $ 98.5\pm0.1  $   \\
   77.1 & $0.14 $fixed     &  $0.fix $  & $ 99.2\pm0.1  $   \\
   71.6 & $0.1 $fixed      &  $0.fix $  & $ 99.5\pm0.1  $    \\
   61.1 & $-0.07\pm0.05$  &  $ 0.fix          $  &  $   98.9\pm0.8  $    \\
   61.2 & $  0.1$fixed     &  $ 1.66 \pm0.5    $  &  $  100.\pm0.2   $     \\
  60.6  & $ 0.0$ fixed     &  $ 0.82 \pm0.5   $  &  $  99.8\pm0.2   $    \\
  60.6  & $ 0.01 \pm0.1$   &  $ 0.74 \pm0.8   $  &  $  99.7\pm0.8   $    \\
\hline
\end{tabular}}
\caption{The basic parameters of the model are determined by fitting experimental data with free $\sigma_{tot}$.}
\label{tab:limits}
\end{wraptable}
However, the size of $\sigma_{tot}$ is the same as in the previous case 
         in the region of  errors.
         If we take the additional part of the slope (the rows 5-7 of  Table 1),
          the error of $\rho$ increases and its size is badly determined.
         But the size of the coefficient $C$ is determined especially with the fixed size of $\rho$.
         In this case, the size of $\sigma_{tot}$ increases  with middle   value $99.7 \ $mb but,
         of course, with a large error.

In Fig. 1a,  the dependence of the size of $\sigma_{tot}$ over $\rho$ (left picture)
 in the case without and with the contributions of the Coulomb and Coulomb-hadron
     interference terms  is shown. The inclusion of the Coulomb dependence terms leads to an increase of $\sigma_{tot}$
     at large $\rho$ and decrease in the case of small and negative $\rho$.

Now let us fix  additional normalization of the experimental data by the middle value
 $1/n$, with $n=1.05$. In this case, we can take the size of  $\sigma_{tot}$ as a
  free parameter. The results are shown in Table 2.
    If $C=0$ (first 5 rows), the size of $\sigma_{tot}$ is less than the determined by the TOTEM Collaboration.
    Again, we can see that if we take $C=0$
  and free $\rho$ and $\sigma_{tot}$, we obtain the negative size of $\rho$. If we take as free parameters
  $\rho$, $C$ and $\sigma_{tot}$, we obtain  $\rho$ near zero.


    Let us check up our assumptions about the size and 
    $t$ dependence of the real part of the scattering
    amplitude. We can use the method which was proposed and explored in  \cite{Sel-Bl95,Sel-PL05}. 
    Let us introduce the value
 \begin{eqnarray}
  \Delta_R(t)=(ReF_C(t)+ReF_h(s,t))^2 = [\frac{d\sigma}{dt}|_{exp.}/n - k \pi *(ImF_c + ImF_h)^2]/(k \pi).
\label{Drds}
  \end{eqnarray}
    For the $pp$ high energy scattering the real part is positive at small 
    $t$ and the Coulomb amplitude is  negative. Hence, the $\Delta_R$ will have the minimum at some point of $t$ and then a wide maximum.

\begin{wraptable}{r}{0.55\textwidth}
\centerline{\begin{tabular}{|c|c|c|c|}
\hline
 $\sum   \chi^{2}_{i}$ & $\rho$  & $C$ & $\sigma_{tot}$, mb  \\\hline
  & & & \\
   77.84 & $  0.14 $fixed    & $ 0. - fix       $     & $   96.8 \pm0.1  $   \\
   71.65 & $  0.1f fix  $    & $ 0.fix             $  & $   97.1\pm0.1   $    \\
   66.3 & $  0.05 fix     $  & $ 0.fix             $  & $   97.2\pm0.1   $    \\
   62.8 & $  0. fix      $   & $ 0. fix           $   & $   97.1 \pm0.1  $     \\
   61.1  & $-0.06 \pm0.05 $  & $ 0. fix          $  & $   96.6\pm0.6   $    \\
   63.1 & $  0.14 $fixed     & $ 2.1\pm0.5     $  & $   97.6 \pm0.2  $   \\
   61.9 & $  0.1fix       $  & $ 1.87\pm0.5    $  & $   97.7\pm0.2   $    \\
   61.0 & $  0.05fix     $   & $ 1.24\pm0.5    $  & $   97.7\pm0.2   $    \\
   60.6 & $  0. fix      $   & $ 0.8\pm0.5     $  & $   97.4 \pm0.2  $     \\
   60.8 & $ -0.05 fix $      & $ 0.4\pm0.5           $  & $   96.9 \pm0.3  $    \\
  60.6  & $-0.01 \pm0.09 $  & $ 0.7 \pm0.9  $  & $   97.3\pm0.9   $    \\
\hline
\end{tabular}}
\caption{The basic parameters of the model are determined by fitting experimental data with $n=1.05$ }
\label{tab:limits}
\end{wraptable}
 Let us take the parameters obtained by the TOTEM Collaboration $\sigma_{tot}=98.6$ mb, $B=19.9 \ $ GeV$^{-2}$, $n=1$,
    $\rho(0)=0.14$ and calculate the value $\Delta_R$, using the left part of eq.({\ref{Drds}). The result is shown in Fig.1b by the hard line.
Now let us take these parameters for the imaginary part 
     and calculate  $\Delta_R$ using the experimental data in the right part of eq.({\ref{Drds}). The triangles 
      in Fig.1b present these calculations. 
      The first and second calculations are very far from each other. If we take the real part with the parameters
    $\sigma_{tot}=96.4$ mb, $B=19.9 \ $ GeV$^{-2}$, $\rho=0.1$ and calculate $\Delta_R$   (short  dashed line in Fig.1b),
    the position of the minimum moves to a higher $t$, but the difference remains 
    large.
 If we take some other parameters for the imaginary part: $\sigma_{tot}=96.4$ mb, $B=20.3 \ $ GeV$^{-2}$,
        $n=1.08$, $C=-0.05$ GeV$^{-1}$, we obtained from  eq.({\ref{Drds}) the result which is shown in Fig.1b by circles.
    The contribution of only the Coulomb amplitude ($\rho=0$) in $\Delta_R$ is shown in Fig.1b
    by the long dashed line and the dotted line represents the calculation with $\rho=-0.05$. We can see that only
    in the last case the difference between the calculations by  eq.({\ref{Drds})(left part)
     and by eq.({\ref{Drds})(right part)
    is not large.

   The analysis of the new experimental data obtained by the LHC TOTEM Collaboration \cite{TOTEM-11a}
   shows that there are some additional specific moments which are to be  taken into account to determine the size of
    $\sigma_{tot}$.  We cannot neglect the electromagnetic interactions.
It is needed to check out the obtained, during the fitting procedure, real part of the scattering amplitude
   by using eq.({\ref{Drds}).
   Maybe, there is some problem with the
    normalization of the separate parts of the experimental data, or there exists some additional (probably oscillation)
    term (see \cite{Osc}) which changes the form of the imaginary part.
     Finally, we should note that the best way to decrease the impact of the different assumptions, which are examined
     in the phenomenological model, consists in the  determination of
    the sizes of $\sigma_{tot}$ and $\rho(s,t)$ simultaneously in one experiment.

\label{sec:figures}
\begin{figure}[htb]
\vspace{-1cm}
\includegraphics[width=0.5\textwidth] {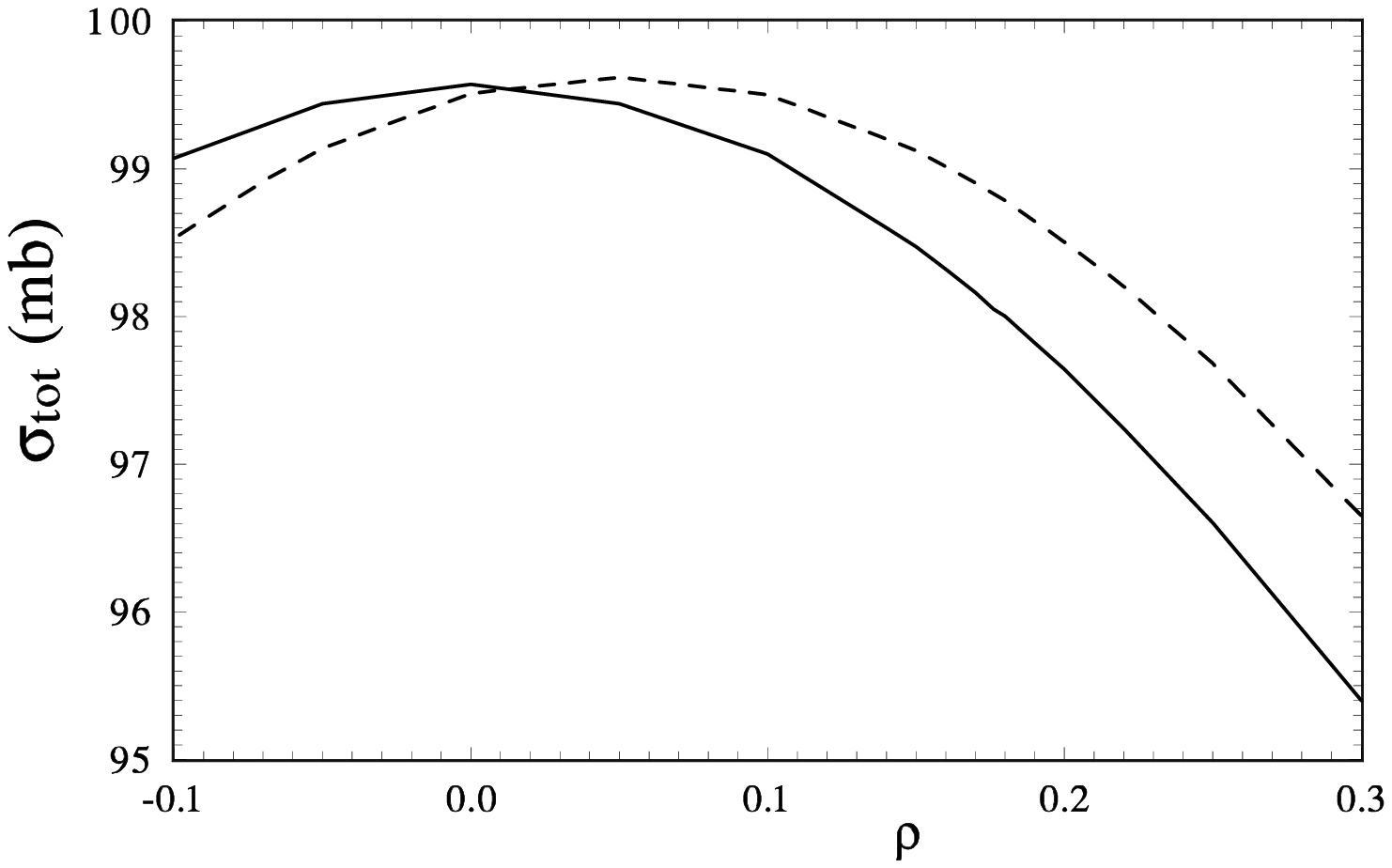}       
\includegraphics[width=0.5\textwidth] {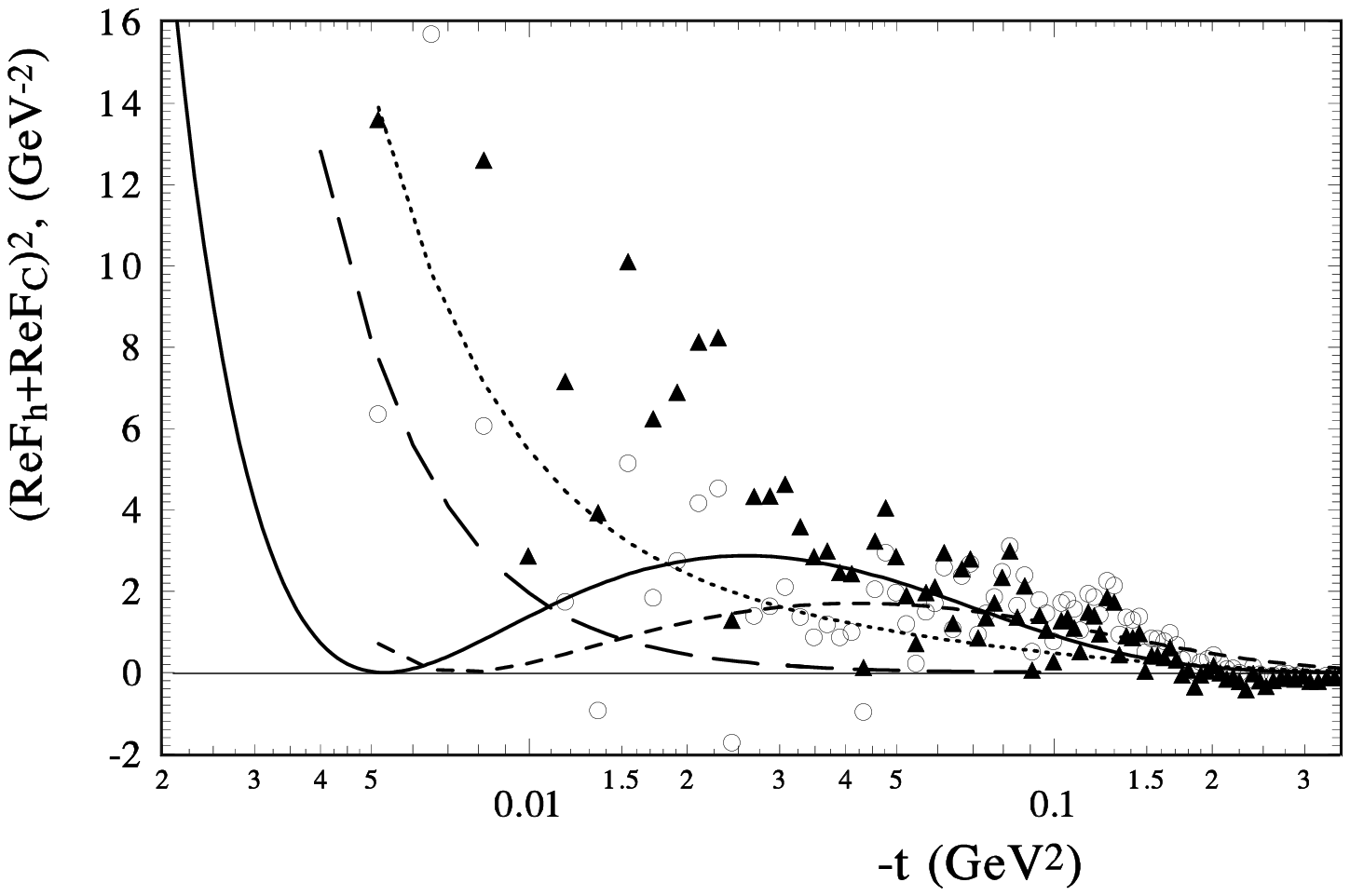}       
\caption{ Size of $\sigma_{tot}$ a)(left)  over $\rho$;  b) (right)
 the calculations of $\Delta_R$.
}\label{Fig:1}
\end{figure}

{\bf Acknowledgments:}  I gratefully acknowledge the organizing committee and R. Orava for the invitation on the conference
  and the financial support. 
  This work is partially supported by WP8 of the hadron physics program of
the 8th EU program period.



\begin{footnotesize}

\end{footnotesize}
\end{document}